\documentclass[useAMS,usenatbib,fleqn]{mn2e}

\usepackage{graphicx}
\usepackage{times}

\usepackage[fleqn]{amsmath} %Gain access to align env. for equation typesetting
\long\def\symbolfootnote[#1]#2{\begingroup%
\def\thefootnote{\fnsymbol{footnote}}\footnote[#1]{#2}\endgroup} 
\usepackage{ifthen}

\newcommand{\dd}{ {\rm d} }                                %Roman Differential d

\newcommand{\ddv}[3][1]{                                   %Differential Fraction with
            \ifthenelse{\equal{#1}{1}}                     %option to show nth derivative as
                       {\frac{\dd #2}{\dd #3}}             %optional argument e.g \ddv[2]{f}{x}
                       {\frac{\dd^{#1} #2}{\dd #3^{#1}}}}  %is d^2 f/ dx^2

\newcommand{\pddv}[3][1]{                                  %As above but with partial differentials
            \ifthenelse{\equal{#1}{1}}
                       {\frac{\partial #2}{\partial #3}}
                       {\frac{\partial^{#1} #2}{\partial #3^{#1}}}}

\newcommand{\kpc}{\ensuremath{\rm kpc}}

\newcommand{\msun}{\ensuremath{\rm M_\odot}}

\newcommand{\Gyr}{\ensuremath{\rm Gyr}}

\DeclareMathOperator{\sech}{sech}

\title[Discy dwarf disruption]
        {Discy dwarf disruption and the shape of the Galactic halo}
\author[Gibbons, Belokurov, Erkal \& Evans]
   {S.L.J. Gibbons$^1$\thanks{E-mail:~sljg2@ast.cam.ac.uk},
    V. Belokurov$^1$, D. Erkal$^1$ and N.W. Evans$^1$
    \medskip
  \\$^1$Institute of Astronomy, University of Cambridge, Madingley Road,
       Cambridge, CB3 0HA, UK}
\begin{document}

\date{Accepted 2016 January 27. Received 2016 January 7; in original form 2015 November 4\\ Accepted version. This is a pre-copyedited, author-produced version of an article accepted for publication in MNRAS
following peer review.}

\pagerange{\pageref{firstpage}--\pageref{lastpage}} \pubyear{2015}

\maketitle

\voffset-1.cm

\label{firstpage}

\begin{abstract}
The shape of the Galactic dark halo can, in principle, be inferred
through modelling of stellar tidal streams in the Milky Way halo.  The
brightest and the longest of these, the Sagittarius stream, reaches
out to large Galactocentric distances and hence can deliver the
tightest constraints on the Galaxy's potential. In this contribution,
we revisit the idea that the Sagittarius Stream was formed from a
rotating progenitor. We demonstrate that the angle between the disk's
angular momentum and the progenitor's orbital angular momentum does
not remain constant throughout the disruption. Instead, it undergoes a
dramatic evolution caused, in part, by the changes in the progenitor's
moment of inertia tensor. We show that, even in a spherical potential,
the streams produced as a result of a disky dwarf disruption appear to be
``precessing''. Yet, this debris plane evolution is illusory as it
is solely caused by the swaying and wobbling of the progenitor's
disk. Stream plane precession is therefore not an unambiguous
indicator of asphericity of the dark halo.
\end{abstract}

\begin{keywords}
Galaxy: halo -- Galaxy: kinematics and dynamics -- galaxies: dwarf:
Sagittarius
\end{keywords}

\section{Introduction}\label{sec:introduction}

The shape and the orientation (if it is aspherical) of the dark halo
of the Milky Way encodes information as to the assembly of the Galaxy,
i.e. its accretion history \citep[e.g.][]{allgood2006}; the
interplay between Dark Matter (DM) and baryons \citep[see
  e.g.][]{kaza2004,victor2008,abadi2010}; the nature of the
DM particle \citep[e.g.][]{dc1991,sidm,mayer2002}; and even the
Cosmology itself \citep[e.g.][]{maccio_cosm}. Complicating the picture
is the fact that the DM halo shape is likely to vary with
Galactocentric radius
\citep[e.g.][]{hayashi2007,veraciro2011}. Unfortunately, only a
handful of techniques currently exist to gauge the Galactic DM halo shape
\citep[e.g.][]{gnedin2005,yu2007,Smith2009,loebman2012}. Of the few
methods available, inference based on the stellar streams dynamics
appears to be one of the most promising methods~\citep[see e.g.][]{helmi_halo, Law2005, Koposov2010}. To probe
regions of the Milky Way dominated by the DM halo, streams reaching to
large Galactocentric radii are required. Covering distances between 15
and 100 kpc, the Sagittarius (Sgr) stream offers exceptional
leverage. However, all attempts to interpret the Sgr stream data thus far
have led to conflicting claims as to the shape of the Galactic halo
\citep[][]{helmi_halo,kvj_halo,fellhauer2006,LM2010}.

The asphericity of the halo produces a non-radial component of the
gravitational force, and the ensuing torques cause the debris plane to
precess. The evolution of the Sgr stream plane has been measured with
high accuracy \citep[][]{Belokurov2014}, but the picture is
confused by the presence of two distinct branches, separated by
$\sim10^{\circ}$, for both the leading and the trailing tail
\citep[][]{field-of-streams,SGB-trailing-paper}. Moreover, as shown by
\citet{Belokurov2014}, the two branches in the bifurcation appear to
precess by different amounts and in different directions. To advance our
understanding of the Galactic gravitational potential, a reliable
model of the Sgr disruption needs to be found that would explain the entirety
of the Sgr data, including the enigmatic stream bifurcation.
\citet{sgrdisk} have shown that a rotating progenitor (i.e. a disk
galaxy) provides a natural explanation of such stream forking. In
their picture, the velocities of unbound stars are a combination of the
progenitor's systemic velocity and the disk rotation velocity at the
moment of stripping. If the disk and the orbital plane are misaligned,
the two velocity vectors change orientation with respect to each other
as the dwarf orbits the Galaxy. Therefore, in every stripping episode,
the debris are sprayed at a different angle with respect to the
orbital plane \citep[see also][]{nicola_streams}. The angular separation between
individual streamlets is controlled by two parameters: i) the angle
between the progenitor's disk and orbital plane; and ii) the
rotational velocity of the progenitor.

\begin{figure}
\centering
\includegraphics[width=0.45\textwidth]{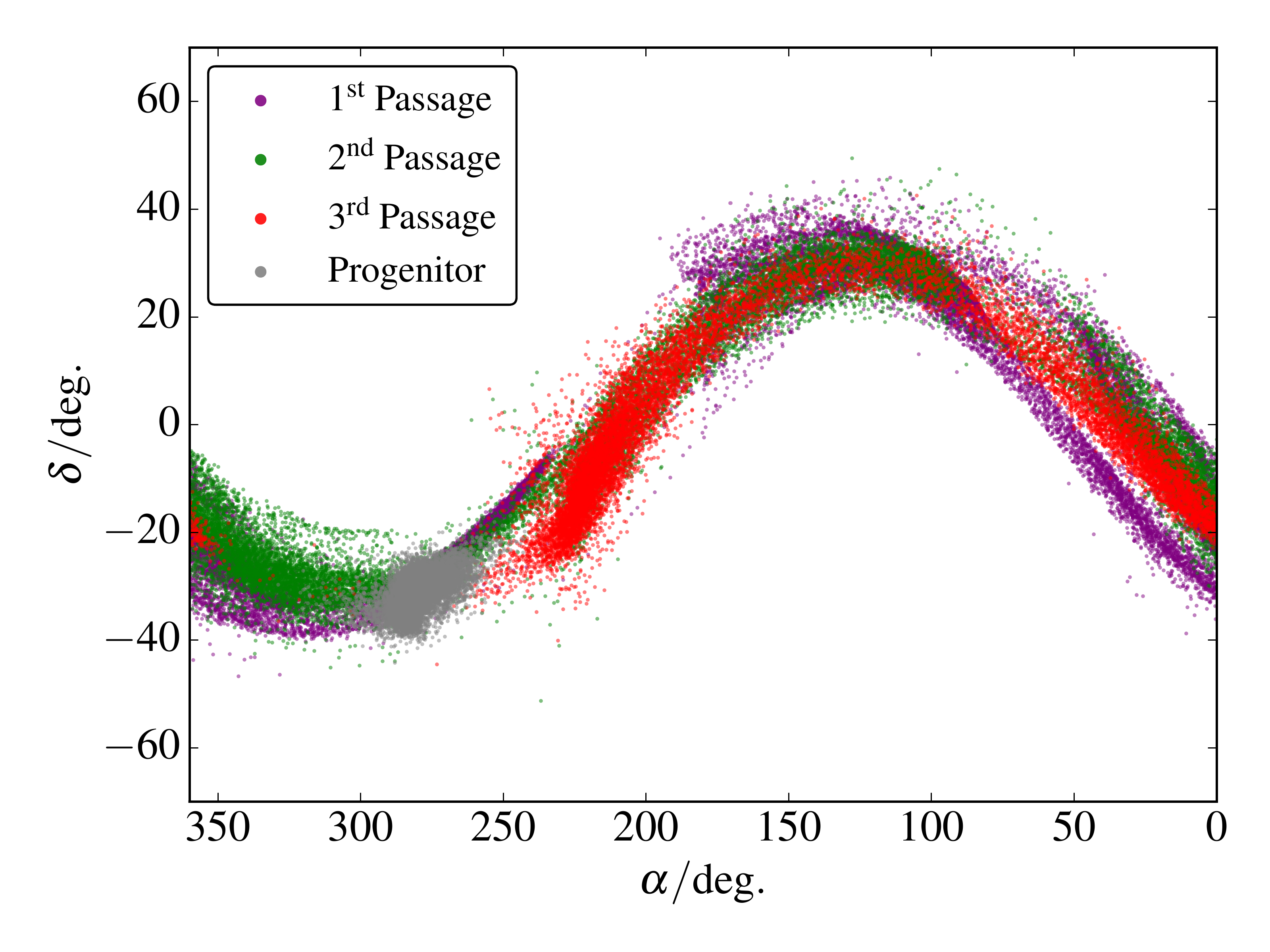}
\caption{Distribution of the stream particles of the leading arm on the sky. The
  particles are color-coded according to the pericentre passage at
  which they are released. In this simulation, the disk's angular
  momentum vector is initially misaligned by $40^\circ$ with the
  orbital angular momentum vector. Note that the stream appearance is
  different from the usual case of a dSph disruption: the streamlets
  corresponding to different pericentre crossing times are sprayed in
  slightly different directions.}
\label{fig:onsky}
\end{figure}
\citet{sgrdisk} predicted that, if their realisation of the Sgr
disruption were correct, a residual rotational signal of $\sim20$
kms$^{-1}$ ought to be present in the remnant. However, a number of
studies have found only a very modest level of rotation in the Sgr
core, typically $< 4$
kms$^{-1}$~\citep[e.g.,][]{jorge2011,frinchaboy2012}. This remains a
serious obstacle for scenarios in which the Sgr progenitor is a
rotating disky dwarf. Even so, the still poorly understood
transformation of rotating dwarf irregulars to non-rotating dwarf
spherodials shows that mechanisms do exist to shed angular momentum.
Alternately, stellar rotation may dominate in the outer parts of the
progenitor, but the inner parts may be supported by velocity
dispersion \citep[e.g., the case of WLM in][]{Le12}, leaving a remnant
with little or no residual rotation after stripping. Whilst
acknowledging the seriousness of this problem, we do believe it is
premature to rule out the entire class of stream models based on
rotating progenitors without a thorough exploration of parameter
space.  Motivated by the development of a new generation of fast and
accurate tidal stream models are becoming available
\citep[see][]{gibbons2014,bowden2015,Kupper2015,amorisco2015}, we
examine a representative simulation of a disky dwarf disruption, in
order to build insight into this largely unexplored regime of stellar
stream formation.

\section{Disky dwarf disruption}

\begin{figure*}
\centering
\includegraphics[width=0.46\textwidth]{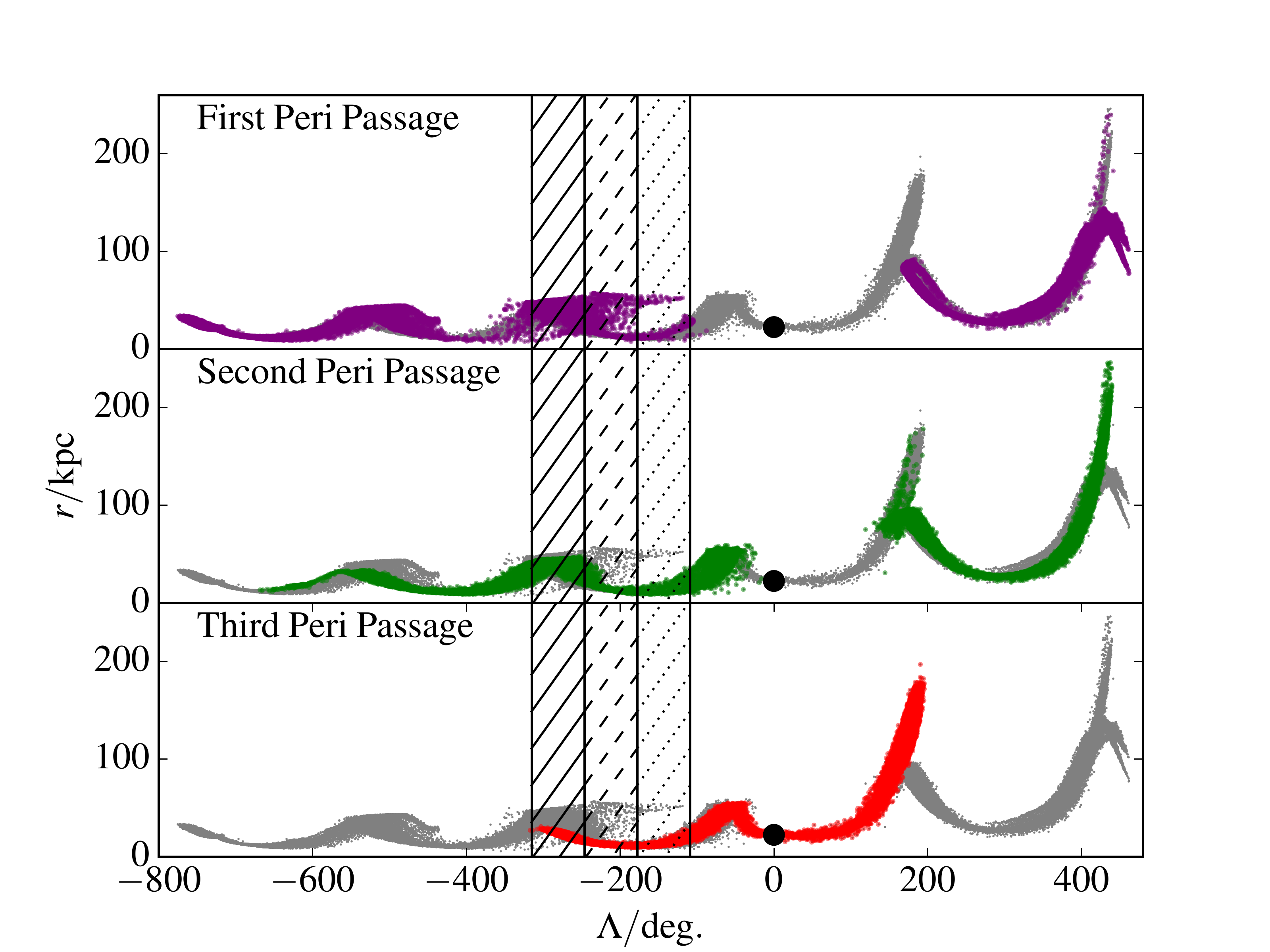}
\includegraphics[width=0.46\textwidth]{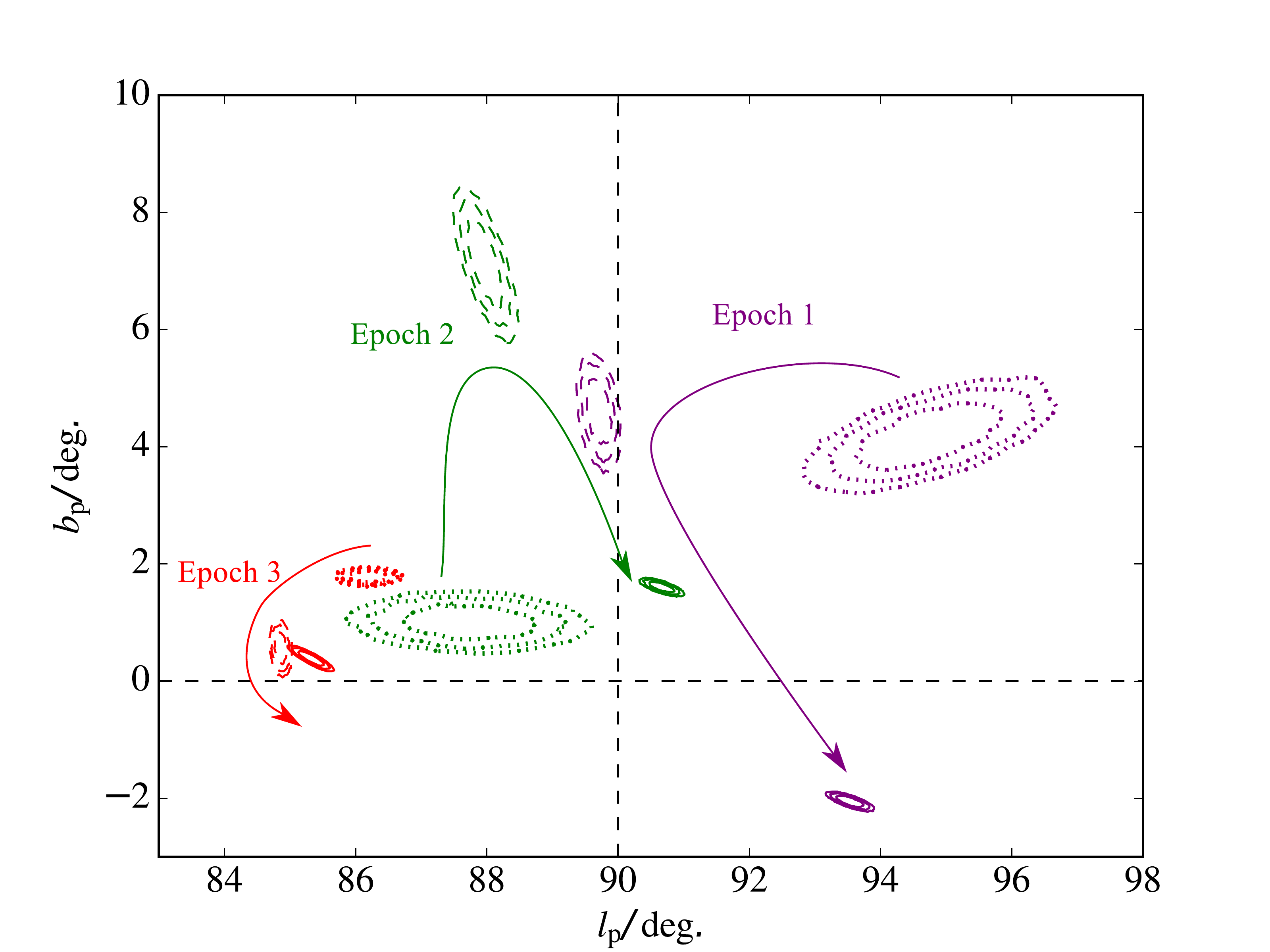}
\label{fig:precession}
\caption{{\it Left:} Simulated stream in ``unwrapped'' coordinates:
  the particles' Galactocentric distance is shown as a function of
  the angle along the stream $\Lambda$. Poles for the debris planes
  are extracted for three $\Lambda$ bins along the leading arm (marked
  by black vertical lines) for each of the three stripping
  epochs. The current position of the progenitor is marked with a black dot.
  {\it Right:} Locations of the poles extracted in the
  Galactocentric coordinates. The dotted, dashed and solid contours
  respectively correspond to the bins marked on the left panel, moving
  away from the progenitor. The colours correspond to the three stripping
  epochs, as defined in the left hand panel.
  The intersection of the black dashed lines indicate the pole of
  the progenitor's orbit. Arrows give the direction of the
  precession away from the progenitor's location. Note that in the
  overlapping regions of the sky, the debris stripped at different
  pericentre passages precess with different amplitudes and in
  different directions.}
\label{fig:poles}
\end{figure*}

To set up the initial conditions of the progenitor we use the {\sc magalie}
code from the {\sc nemo} stellar dynamics toolbox. This
employs the algorithm presented in \citet{magalie} to construct
realisations of composite progenitors comprised of
stellar disks embedded within DM haloes. In choosing the structural
parameters for the progenitor we followed \citet{sgrdisk}. More
precisely, for the stellar disk we used an exponential density profile:
\begin{align}
\rho_{\rm disk} = \frac{M_d}{4\pi R_d^2 z_d} \exp(-R/R_d) \sech^2(z/z_d) , 
\end{align}
\noindent where $M_d$ is the mass, $R_d$ is the radial scale length of
the disk and $z_d$ is the vertical scale height. We chose to fix
these at $M_d=3.5\times 10^8 \msun$, $R_d = 0.9 \kpc$. Additionally,
we assumed that the vertical scale height is related to the radial
scale length by $z_d = 0.2 R_d$. For the progenitor's DM halo, we used
the density profile:
\begin{align}
\rho_{\rm halo} = \frac{M_h \alpha(r_c/r_{\rm cut})}{2\pi^{3/2} r_{\rm cut}} \frac{\exp(-r^2/r_{\rm cut}^2)}{r^2 + r_c^2},
\end{align}
\noindent where $M_h=2.4\times 10^9 \msun$ is the halo mass, $r_c=0.5R_d$ is
the core radius and $r_{\rm cut} = 6 R_d$ is the radius at which the halo is
truncated. For these values of the structural parameters $\alpha \approx 1.156$.

We imposed the truncation to the halo as the outer
envelope of the progenitor's dark halo is likely to be lost before the stream
begins to form.  This is in line with the results reported by
\citet{Martin2010} who found that the outer parts of the halo are lost before
the majority of stellar particles are stripped. The dwarf's circular velocity
curve is therefore a combination of the contributions from the disk and the
halo, and peaks at $47$ kms$^{-1}$ at 2.7 kpc.

We evolved the progenitor in a fixed spherical NFW model
\citep{NFW1996} with $M_{200} = 7.5 \times 10^{11} \msun$ and
concentration $c_{200}=20$. This is not meant to be a realistic model of the
Galactic potential but is just a simple representation, convenient in
isolating the effects of rotation in the progenitor on the stream
particles, aside from any asphericity in the host potential. The
satellite is placed on an orbit with an apocentre of $r_a = 70 \kpc$, a pericentre of $r_p=18\kpc$, and an orbital period of $\approx
1.2 \Gyr$. The simulations were carried out using the NBODY solver in
{\sc Gadget-2} \citep{Springel2005} which was modified to implement a
static NFW potential by adding an additional force component,
dependent on position, at each force computation.

Figure~\ref{fig:onsky} displays the on-sky distribution of particles of the
leading arm for a satellite disruption simulation where the dwarf's disk is
initially misaligned with its orbital plane by $40^{\circ}$. The stream
particles are coloured according to the epoch during which they became unbound
from the progenitor: the first to go were the particles shown in purple (1st
pericentric passage); followed by the green; and finally the most recently
stripped particles are shown in red.  If the stream's progenitor were not
rotating or if the disk and the orbital plane were aligned, given that the
host's potential is spherical the debris stripped at different epochs would
keep piling up on top of one another.  However, as shown in the Figure, the
situation is markedly different for the misaligned disruption, as already found
by \citet{sgrdisk}.  Here, streamlets corresponding to individual stripping
episodes are clearly offset on the sky, implying that they stay in distinct
orbital planes.

\section{Apparent stream precession and disk wobbling}\label{sec:analytics}

{\bf Streamlet precession.} Let us have a closer look at the plane
evolution of the debris stripped during each pericentre crossing. The
left panel of Figure~\ref{fig:poles} gives the ``unwrapped'' view of
the stream in the plane of Galactocentric distance and angle along
the stream $\Lambda$. In these coordinates, it is straightforward to
isolate the debris belonging to each stripping episode. Having thus
separated the particles into three groups, we can measure their orbital
plane evolution. More precisely, for each bin in $\Lambda$ (the
boundaries of the three bins used in this analysis are shown with
black vertical lines in the left panel of the Figure), the plane of
the debris is assumed to be constant, well-described by its normal
vector $\hat{\mathbf{n}}$, with the particles scattered about this
plane by a Gaussian with variance $\Delta$. We chose to
parameterize the normal vector in terms of Galactocentric coordinates
defined as $\hat{\mathbf{n}} = \cos b_{\rm p}\cos l_{\rm
  p}\hat{\mathbf{x}} + \cos b_{\rm p}\sin l_{\rm p}~\hat{\mathbf{y}}
+ \sin b_{\rm p}~\hat{\mathbf{z}}$

With these assumptions, the likelihood of the given normal and the
variance is thus:
\begin{align}
\log \mathcal{L} = - \sum_{i=1}^N \frac{(\hat{\mathbf{n}} \cdot \mathbf{x}_i)^2}{2\Delta} - \frac{N}{2} \log \Delta,
\end{align}

\noindent where $N$ is the number of particles in the bin, the
$\mathbf{x}_i$'s are the 3D positions of each of the particles and the
sum is taken over all of the particles in the bin.

Debris plane evolution for each of the three stripping epochs is
presented in the right panel of Figure~\ref{fig:poles}. It is clear
that as one steps along the leading arm of the stream, away from the
progenitor (as indicated by varying contour styles and the arrows),
swarms of particles from each of the three pericentric passages show
qualitatively dissimilar behavior, with the plane of each individual
spray of debris precessing with a different amplitude and direction.
This behavior is qualitatively similar to that observed in Figure~13
of \citet{Belokurov2014} where the bright arm of the bifurcation
(Branch A) appears to be precessing with a much smaller amplitude, and
in the opposite direction to, the faint arm of the stream (Branch
B). These results therefore suggest that the bright arm of the
bifurcation could have been formed from the debris that is
(dynamically) younger than the faint arm. Curiously, this is conflicts with the
simulation presented by \citet{sgrdisk}. We note, however, that our
interpretation is in agreement with the recent measurement of the chemical
abundance differences between the two branches by \citet{SGB-trailing-paper}. They
report that Branch B appears to lack the metal-rich populations that are discernible in
Branch A, therefore implying that this debris could have been stripped
earlier.

\begin{figure}
\centering
\includegraphics[width=0.35\textwidth]{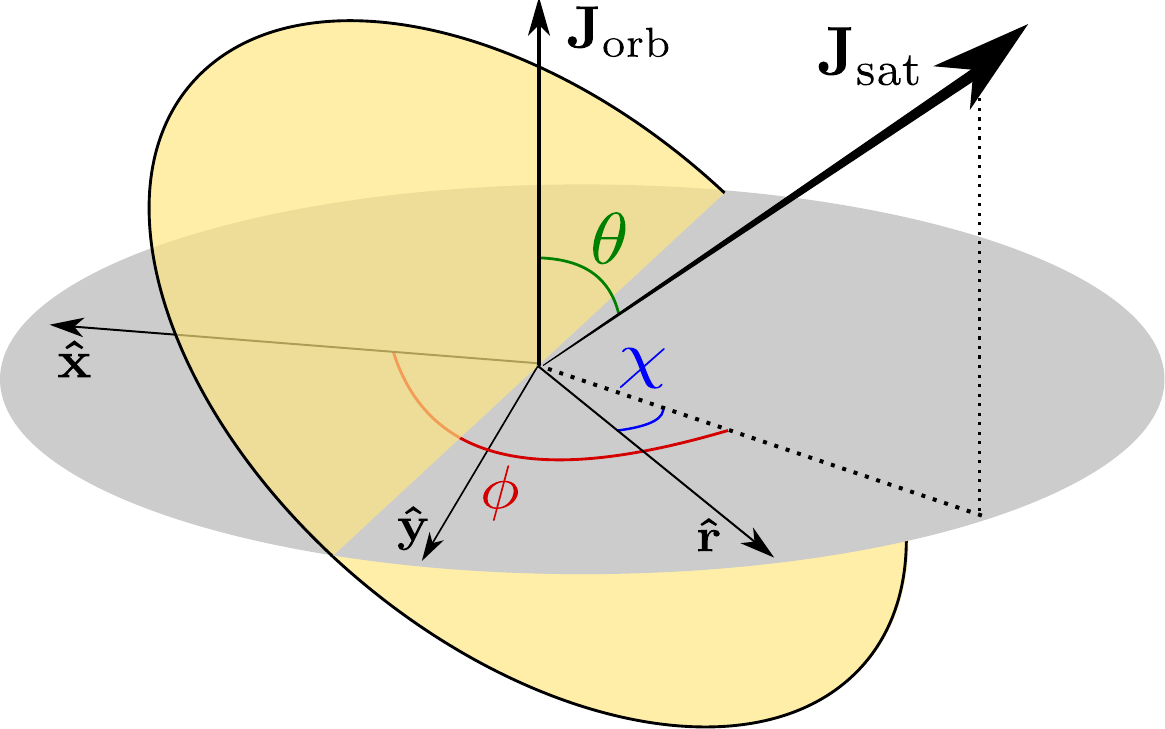}
  \caption{A schematic diagram showing the orientation angles of the
    disk. $\theta$ is the angle between $\mathbf{J}_{\rm orb}$ and
    $\mathbf{J}_{\rm sat}$. $\phi$ and $\chi$ are the angles between
    the projection of $\mathbf{J}_{\rm sat}$ into the plane of the
    orbit and the Galactocentric $\mathbf{\hat{x}}$ and $\mathbf{\hat{r}}$ vectors
    respectively.}
\label{fig:angle_cartoon}
\end{figure}
\begin{figure*}
\centering
\includegraphics[width=0.33\textwidth]{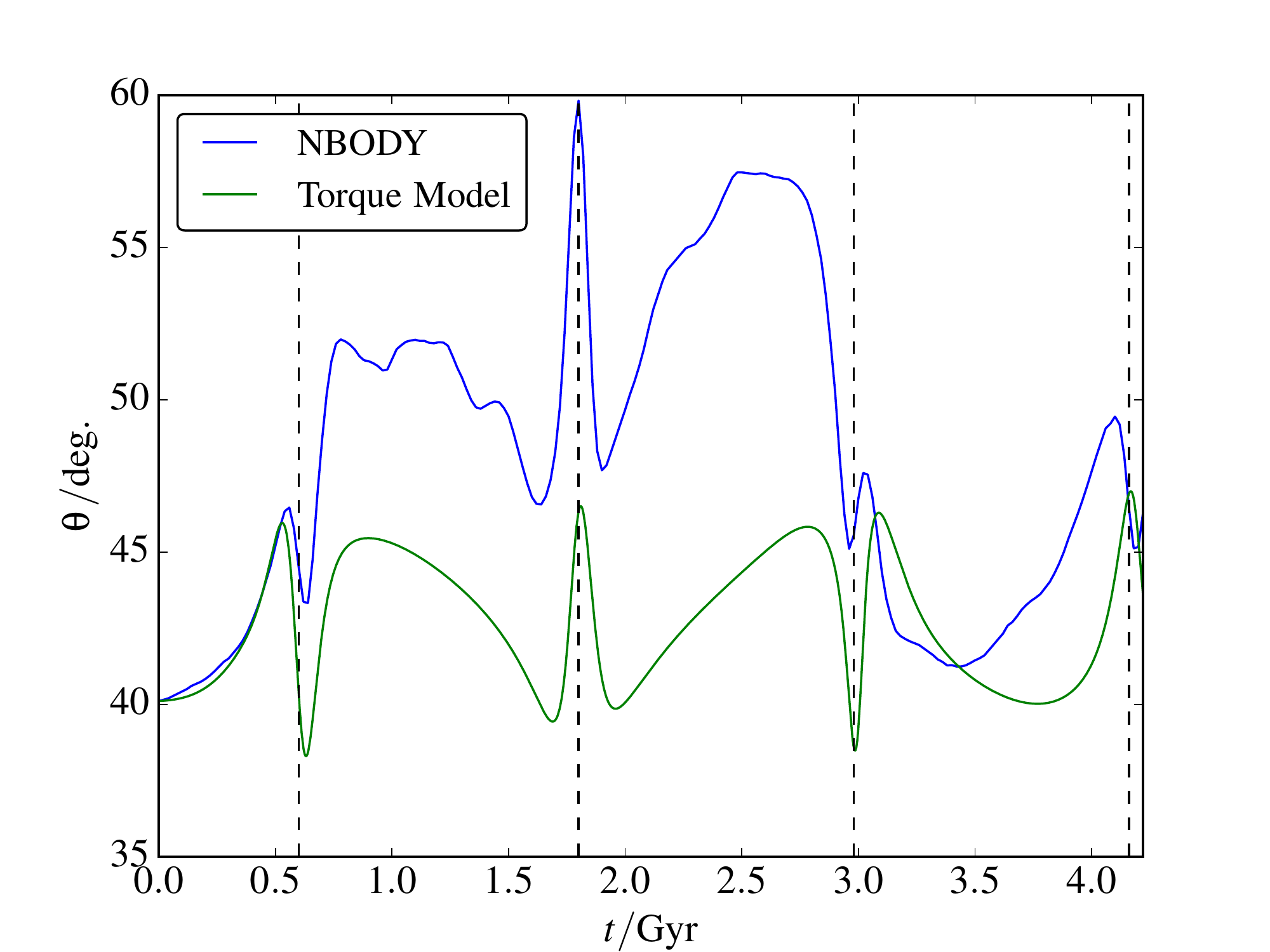}
\includegraphics[width=0.33\textwidth]{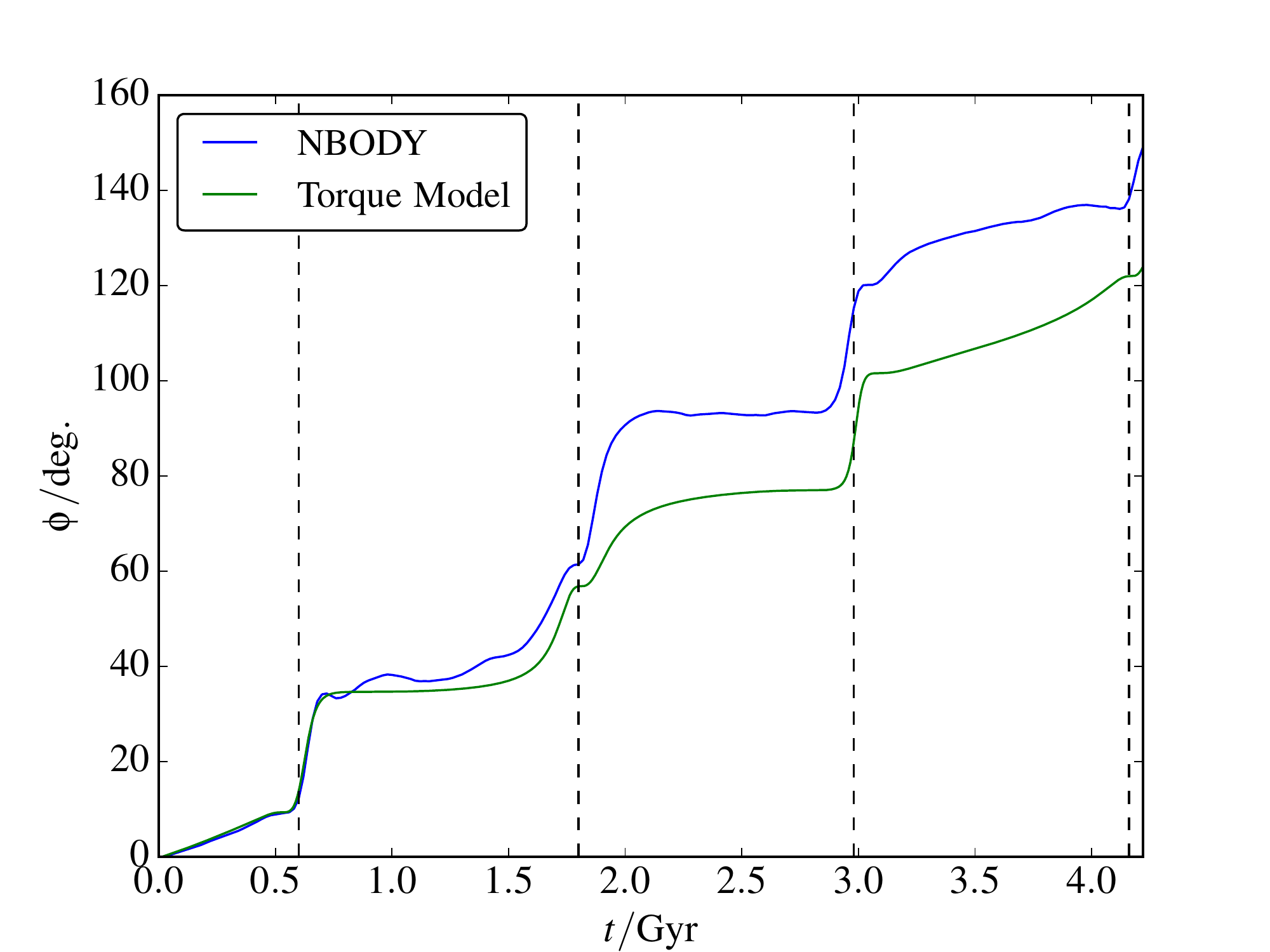}
\includegraphics[width=0.33\textwidth]{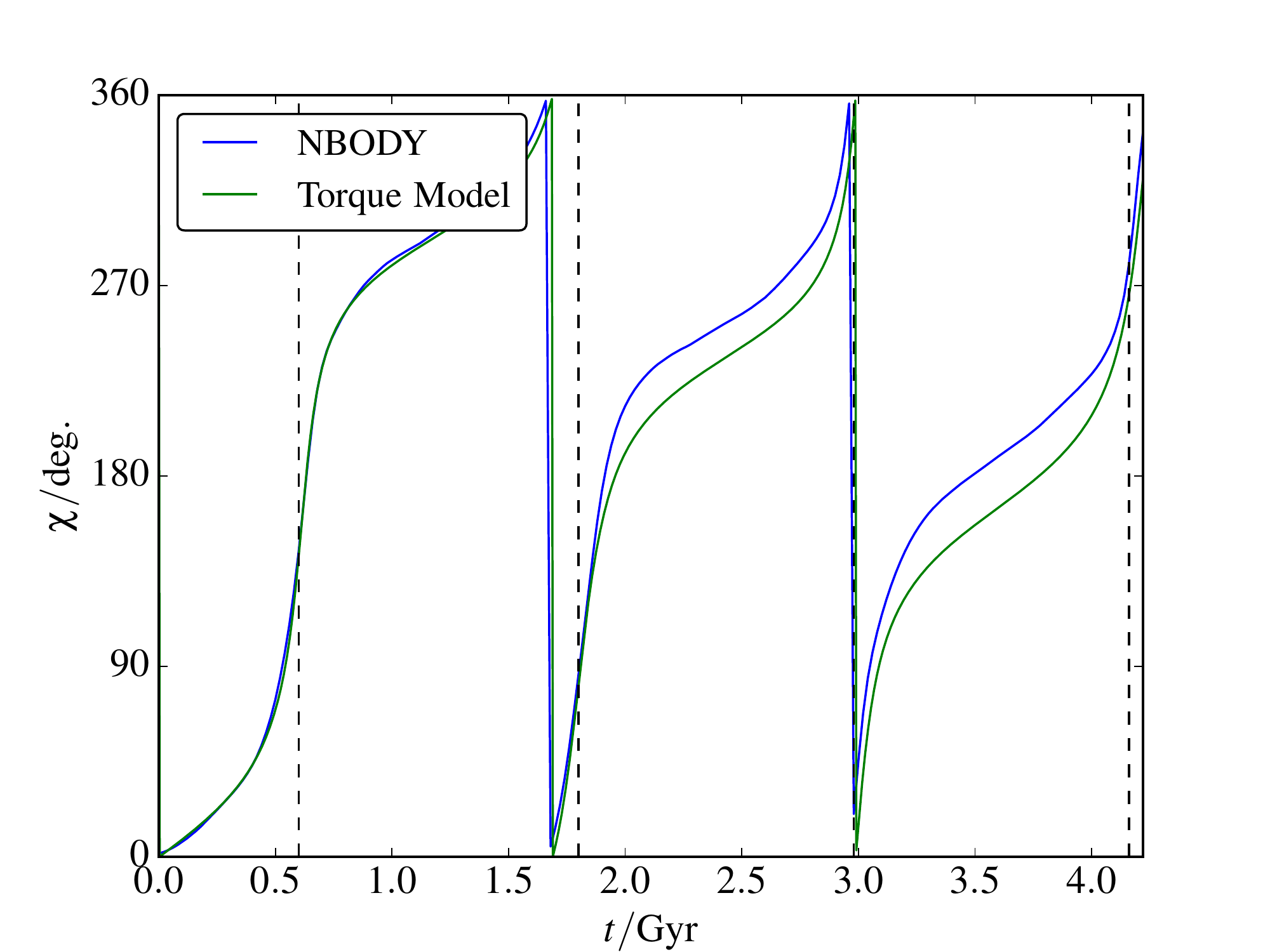}
\caption{Evolution of the three orientation angles of the disk, as
  defined in Figure~\ref{fig:angle_cartoon} in the simulation (blue)
  and as predicted from the formalism presented in
  section~\ref{sec:analytics} (green). The times of the pericentre
  passages are indicated with dashed black lines. The NBODY and torque models
  agree closely until the first pericentre passage, after which the agreement is less
  exact. However there is good qualitative agreement over the entire period
  simulated.}
\label{fig:torque}
\end{figure*}

{\bf Disk wobble and precession.} The strong apparent precession of
the streamlets that is visible in the right panel of Figure~\ref{fig:poles} is
surprising as the simulation of the disky dwarf disruption was
performed in a spherical potential, where there are no torques acting
on the stripped particles. However, there are torques acting on the
progenitor's disk, making it wobble and precess as it disrupts. The
torque acting on a disk, with a density distribution
$\rho(\mathbf{r})$, located in an external potential $\Phi$, and with
its centre of mass at a position $\mathbf{r_0}$ is:

\begin{align}
\frac{\dd \mathbf{J}_{\rm sat}}{\dd t} = -\int \dd^3\mathbf{r} ~ \rho(\mathbf{r}) ~ \mathbf{r} \times \nabla \Phi(\mathbf{r_0 + r}) .
\end{align}

If we assume that the force on the disk is slowly varying with
position, we can expand the gradient of the potential about the center
of mass of the progenitor to obtain:

\begin{align}
\left. \nabla \Phi(\mathbf{r_0 + r})\right|_i \approx \left. \nabla \Phi(\mathbf{r_0})\right|_i + \sum_{k=1}^3 H_{ki} x_k + \mathcal{O}(x^2) .
\end{align}

\noindent Here $H_{ij} \equiv \partial_i \partial_j \Phi$ is the Hessian of
the potential evaluated at the centre of mass of the progenitor. In a frame
centered on the disk with basis vectors $\mathbf{e}_1$, $\mathbf{e}_2$ and
$\mathbf{e}_3$ aligned with the symmetry axes of the disk, the torque is

\begin{align}
\frac{\dd \mathbf{J}_{\rm sat}}{\dd t} \approx H_{23} (I_3 - I_2) \mathbf{e}_1 + H_{13} (I_1 - I_3) \mathbf{e}_2 + H_{12} (I_2 - I_1) \mathbf{e}_3,
\label{eq:dj}
\end{align}

\noindent where the $I_{i}$ are the principal axes of the moment of
inertia tensor of the disk, defined as $I_i = \int \dd^3 \mathbf{r} ~
\rho(\mathbf{r}) x_i^2$.  For an exponential disk, these can be
evaluated as:
\begin{align}
I_1 = I_2 = 3 M_d R_d^2 ~~~ ; ~~~ I_3 =  \frac{1}{12}M_d \pi^2 z_d^2
\end{align}

We followed changes in the disk's orientation during its orbit around
the Milky Way, as described by the following three angles (see
Figure~\ref{fig:angle_cartoon}): $\theta$ is the angle between the
orbital angular momentum, $\mathbf{J}_{\rm orb}$, and the satellite's
internal angular momentum, $\mathbf{J}_{\rm sat}$; $\phi$ is the angle
between the projection of $\mathbf{J}_{\rm sat}$ onto the plane of the
orbit and the Galactocentric $x$ direction; finally, $\chi$ is the
angle between the radial vector $r$ and the projection of
$\mathbf{J}_{\rm sat}$ onto the plane of the
orbit. Figure~\ref{fig:torque} shows the evolution of each of the
three angles defined above. The model specified by
Equation~\ref{eq:dj} (green curve) can be compared with the actual
trajectory of the disk's pole (blue curve), as measured from the
particles in the N-body simulation. The left panel of the Figure
reveals a noticeable change in the misalignment angle between the disk
and the orbital plane $\theta$ as a function of time. Fast dramatic
oscillations in $\theta$ are visible around the times of the
pericentre crossing. As the progenitor approaches the pericentre, the
Galaxy pulls the disk's nearest edge. Once passed, the other side of
the disk is tugged, making it rock back and forth. Specifically, the
sign of $\dd \theta / \dd t$ switches every time the disk's viewing
angle, $\chi$, crosses an integer multiple of $90^{\circ}$, as
illustrated in the right panel of Figure~\ref{fig:torque}. The growth rate of
$\chi$ is primarily governed by the orbital angular velocity of the progenitor
and therefore depends on the eccentricity of the orbit. The orbit of the Sgr
dwarf is rather eccentric and thus, around each pericentre, $\chi$ sweeps
through more than $90^\circ$ in a small fraction of the orbital period, causing
fast oscillations in $\theta$. On top of the periodic wobble seen in the left
panel of the Figure, the Galaxy's torques naturally also cause the disk to
precess. This precession, i.e. the evolution of the angle $\phi$, is shown in
the middle panel of the Figure. Note that $\phi$ always evolves in the same
direction, and over the course of the disruption, the amplitude of its change
is much larger than that of $\theta$.

Therefore we conjecture that: a) the overall difference \emph{between} debris plane
poles of individual streamlets result from the precession of the disk, i.e.
changes in $\phi$;  b) \emph{within} each individual streamlet, the debris pole
changes are caused by the combination of disk wobble and disk
precession, i.e. changes in both $\theta$ and $\phi$.

\begin{figure}
\centering
\includegraphics[width=0.46\textwidth]{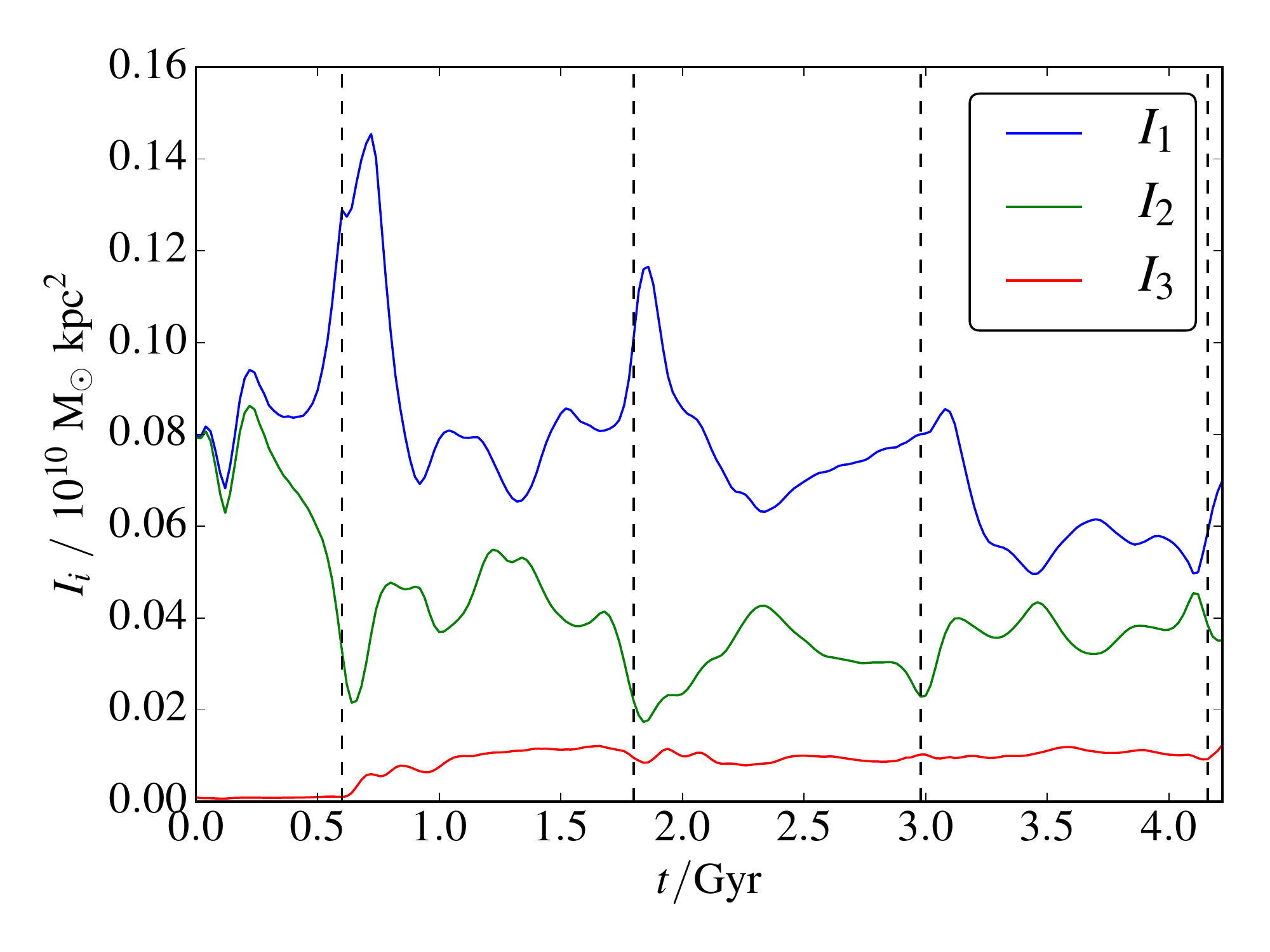}
\caption{Evolution of the three principal axes of the moments of
  inertia tensor of the disky progenitor in the simulation with an initial
  misalignment angle of $40^\circ$. Note, that as material is being
  stripped, the disk is also stirred into a more spherical shape.}
\label{fig:inertia}
\end{figure}

There is a good level of agreement between the model and the simulation: the
formalism described above clearly is able to capture the qualitative
behaviour. However, the actual disk wobble appears to be
following a more complex pattern. We ascribe these differences to the changes
in the satellite's moment of inertia tensor that are not captured by our simple
analytical torque model. Above, $I_{i}$ are assumed to be constant. Yet, as the disk
is disrupted by the Galaxy's tides, its shape undergoes a dramatic evolution,
as illustrated in Figure~\ref{fig:inertia}. As can be gleaned from the Figure,
the disky dwarf is becoming more spherical as it sheds stars. This is indicated
by each of the $I_i$s becoming closer in value as time progresses. This is in
agreement with the earlier work of \citet{Mayer2001} who invoked this phenomenon
to explain the morphological transformation of dwarf galaxies.

\section{Discussion and Conclusions}\label{sec:conclusions}

It is natural to assume that the progenitor of the Sgr stream was most
likely rotating.  The luminous matter in the Sgr dwarf was reckoned by
\citet{Martin2012} to be $\sim 10^8 L_\odot$. This is the typical
luminosity of dwarf irregular or dwarf elliptical galaxies, both of
which possess intrinisic rotation. Therefore, the effects of rotation
in the progenitor are probably important in understanding the
properties of the Sgr tidal debris. Against this must be set the fact
that the present day Sgr remnant is barely rotating at
all~\citep{jorge2011,frinchaboy2012}.

It has long been hoped that precession of the debris plane, as inferred
from the analysis of stellar streams like Sgr, can yield
measurements of the flattening of the gravitational potential of the
Milky Way. Orbits in a spherical potential are confined to a
plane. Once the potential is made aspherical, then the plane begins to
precess with a rate that depends on the flattening of the potential.
However, as we have shown here, even in a spherical potential, the debris
plane of a stellar stream might appear strongly to be ``precessing'' if it is
produced by the disruption of a progenitor with a rotating disk, whose
rotation is misaligned with the orbital plane.

We have established that the evolution of the debris plane along the
stream (and along the individual streamlets, or branches) can also be
caused by the disk wobbling and precessing during the pericentre
crossing.  We have developed a simple model of the disk orientation
and the corresponding stream ``precession'', which captures the
qualitative behaviour. Through tidal stirring, the progenitor's moment
of inertia tensor evolves away from its original configuration and so
our simple model breaks down at large times. Thus, it is not yet
possible to extend the fast and accurate stream models, similar to
those discussed in \citet{gibbons2014}, to include rotating
progenitors. Nevertheless, we find an encouraging level of agreement in
the debris plane behaviour between our simulation and the Sgr stream
measurements presented by \citet{Belokurov2014}. We conclude that it
is premature to discard the disky dwarf hypothesis of \citet{sgrdisk}
and caution against direct interpretation of the stream plane
precession as caused by a non-spherical DM halo.

\section*{Acknowledgements}
The authors thank the anonymous referee for a constructive report. SG thanks
Michael Alexander for a thorough proofreading of the manuscript and
acknowledges the Science and Technology Facilities Council (STFC) for the award
of a studentship. The research leading to these results has received funding
from the European Research Council under the European Union's Seventh Framework
Programme (FP/2007-2013) / ERC Grant Agreement no. 308024.

\label{lastpage}

\footnotesize{
    \bibliographystyle{mn2e}
    \bibliography{refs}
}

\end{document}